\documentclass[a4paper,useAMS,usenatbib]{mnras}

\usepackage{graphicx}
\usepackage{setspace}
\usepackage{natbib}
\usepackage{color}
\usepackage{amsmath,amssymb}
\usepackage{times}
\usepackage{hyperref}

\title[The Pisces Plume]{The Pisces Plume and the Magellanic Wake}

\author[Belokurov et al]{Vasily
  Belokurov$^{1,2}$\thanks{E-mail:vasily@ast.cam.ac.uk}, Alis
  J. Deason$^{3}$, Denis Erkal$^{4}$, Sergey E. Koposov$^{1,5}$,
\newauthor Julio~A. Carballo-Bello$^{6,7}$, Martin C Smith$^8$, Prashin Jethwa$^{9}$, Camila Navarrete$^{6,10}$\\
  $^{1}$Institute of Astronomy, Madingley Rd, Cambridge, CB3 0HA\\
$^{2}$Center for Computational Astrophysics, Flatiron Institute, 162 5th Avenue, New York, NY 10010, USA\\
$^{3}$Institute for Computational Cosmology, Department of
  Physics, University of Durham, South Road, Durham DH1 3LE, UK\\
  $^{4}$Department of Physics, University of Surrey, Guildford GU2 7XH, UK\\
  $^5$Department of Physics, McWilliams Center for Cosmology, Carnegie Mellon University, 5000 Forbes Avenue, Pittsburgh, PA 15213, USA\\
  $^{6}$Instituto de Astrof\'isica, Pontificia Universidad Cat\'olica de Chile, 
Av. Vicu\~na Mackenna 4860, 782-0436 Macul, Santiago, Chile \\ 
$^{7}$Chinese Academy of Sciences South America  Center for Astronomy, National Astronomical Observatories, CAS, Beijing 10010\\
  $^8$Key Laboratory for Research in Galaxies and Cosmology, Shanghai Astronomical Observatory, Chinese Academy of Sciences, 80 Nandan Road, \\Shanghai 200030, People’s Republic of China\\
$^{9}$European Southern Observatory, Karl-Schwarzschild-Str. 2, 85748 Garching, Germany\\
$^{10}$10 Millennium Institute of Astrophysics, Av. Vicu{\~n}a Mackenna 4860,
782-0436, Macul, Santiago, Chile
}

\begin{document}

\defcitealias{H16}{H16}


\maketitle

\label{firstpage}

\begin{abstract}
Using RR Lyrae stars in the Gaia Data Release 2 and Pan-STARRS1 we
study the properties of the Pisces Overdensity, a diffuse
sub-structure in the outer halo of the Milky Way. We show that along
the line of sight, Pisces appears as a broad and long plume of stars
stretching from $40$ to $110$ kpc with a steep distance gradient. On
the sky Pisces's elongated shape is aligned with the Magellanic
Stream. Using follow-up VLT FORS2 spectroscopy, we have measured the
velocity distribution of the Pisces candidate member stars and have
shown it to be as broad as that of the Galactic halo but offset to
negative velocities. Using a suite of numerical simulations, we
demonstrate that the structure has many properties in common with the
predicted behaviour of the Magellanic wake, i.e. the Galactic halo
overdensity induced by the in-fall of the Magellanic Clouds.

\end{abstract}

\begin{keywords}
Magellanic Clouds -- galaxies: dwarf -- galaxies: structure -- Local Group -- stars: variables: RR Lyrae
\end{keywords}

\section{Introduction}

The implications of the arrival of a massive \citep[see
  e.g.][]{Jorge2016} Large Magellanic Cloud (LMC) on a highly
eccentric and rapidly evolving orbit \citep[][]{Besla2007,
  Kallivayalil13} are many fold. First, in response to the interaction
with a massive satellite, the barycenter of the Milky Way should move
\citep[][]{Gomez2015}, imprinting noticeable velocity gradients
throughout the Galaxy \citep[][]{orphan_lmc_mass}. Second, the MC
fly-by will violently disturb the Galactic disc (both stellar and
gaseous), warping and heating it out of equilibrium
\citep[][]{Laporte2018}. Finally, the giant satellite will
gravitationally focus dark matter particles directly behind it, as it
sinks to the centre of the Milky Way due to dynamical friction
\citep[][]{lmc_wake}. This large-scale ``wake'' ought to have an
observable stellar counterpart, composed of MW halo stars at distances
beyond $\sim40$ kpc.

\begin{figure*}
  \centering
  \includegraphics[width=0.64\textwidth]{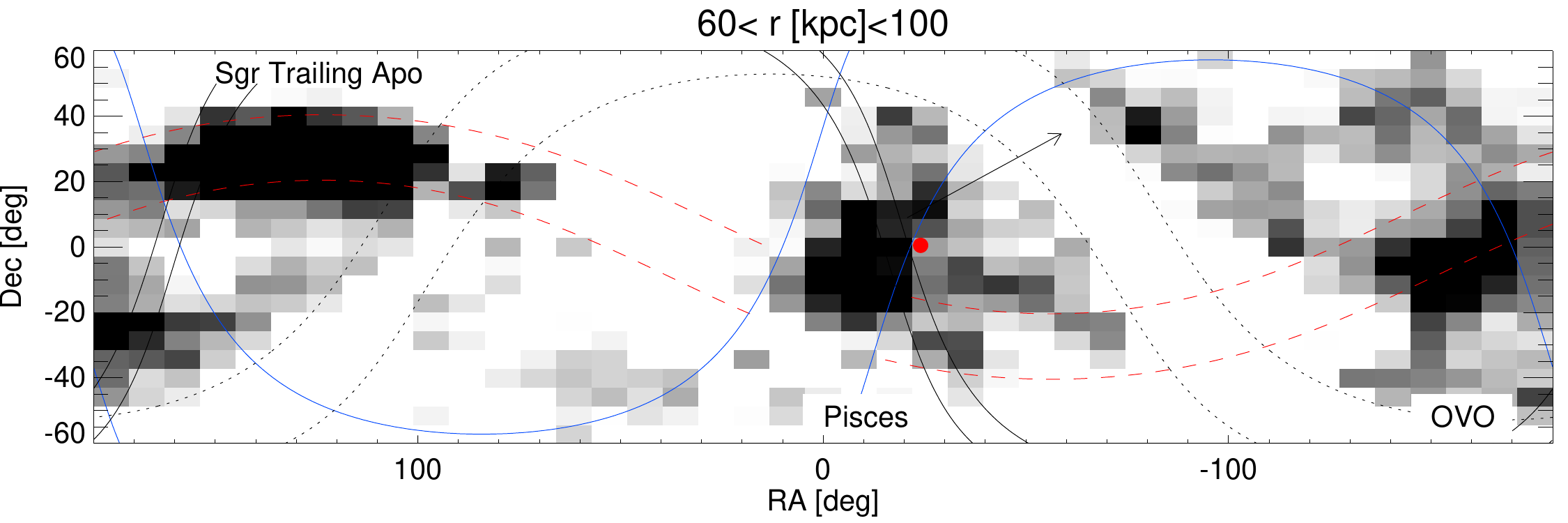}
  \includegraphics[width=0.313\textwidth]{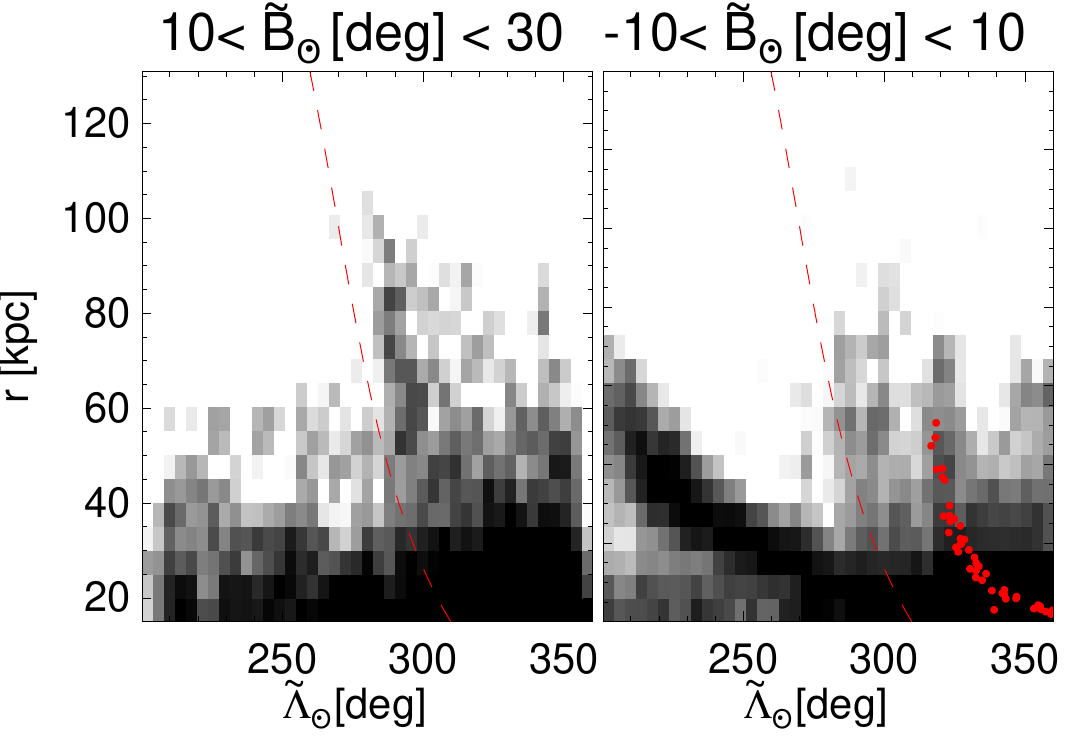}
  \caption[]{{\it Left:} Density of distant GDR2+PS1 RR Lyrae
    candidates in equatorial coordinates. Stars as close as $\sim$60
    kpc and as far as $\sim$100 kpc are included. Three obvious
    over-densities are visible: the apo-centric pile-up of the Sgr
    dsph trailing debris at $90^{\circ}<{\rm RA}<160^{\circ}$, the
    Outer Virgo Over-density (OVO) at $-180^{\circ}<RA<-140^{\circ}$
    and the Pisces cloud of stars at
    $-30^{\circ}<$RA$<10^{\circ}$. The solid blue lines indicate the
    region within the Magellanic Stream, i.e. $-15^{\circ}<B_{\rm
      MS}<15^{\circ}$. The red dashed lines give the Sgr stream track,
    i.e. $-10^{\circ}<\tilde B_{\odot} < 10^{\circ}$. The solid black
    lines are offset $\pm3^{\circ}$ from the Orphan Stream track. The
    Galactic plane ($b=\pm10^{\circ}$) is marked with dotted black
    lines. The black arrow is the direction of the Pisces elongation
    as measured by \citet{Nie2015}. The filled red circle marks the
    location of the HSC's VVDS field. {\it Right:} Heliocentric
    distance distribution of the GDR2+PS1 RR Lyrae as a function of
    Sgr's longitude $\tilde{\Lambda}_{\odot}$ for two different ranges
    of Sgr's latitude $\tilde{B}_{\odot}$. The Pisces distance
    gradient estimated by \citet{Nie2015} using SCUSS BHBs is shown as
    a red dashed line (offset by $-15^{\circ}$ from their central
    location). Small red filled circles are the OS RR Lyrae from \protect\cite{Koposov2019} with
    $\tilde B_{\odot} < 10^{\circ}$.}
   \label{fig:distant}
\end{figure*}

In this Letter, we explore the structure of the Galactic stellar halo
at distances beyond $\sim40$ kpc from the Sun using a large sample of
RR Lyrae provided by the Gaia Data Release 2
\citep[][]{Clementini2018,Holl2018} as well as the Pan-STARRS1 survey
\citep[PS1,][]{Sesar2017}. We show that the structure known as the
Pisces Overdensity is in fact a part of a long, broad and nearly
radial stream, which extends from $\sim40$kpc to $>100$kpc in
distance. Pisces was discovered as an over-density of RR Lyrae stars
with heliocentric distances $\sim$80 kpc in the SDSS Stripe 82 data
\citep[][]{Sesar2007,Watkins2009}, which provided only a limited view
of the cloud. Most recently, \citet{Nie2015} used a large sample of
photometrically selected Blue Horizontal Branch (BHB) stars to map out
the full extent of the Pisces structure. This expanded view revealed
that rather than a cloud, the Pisces over-density i) shows a steep
distance gradient and ii) has a clearly elongated shape on the
sky. Here we revisit the position and the highly stretched 3D
structure of Pisces. Additionally, we collect and analyze kinematical
information along a sightline through the Overdensity. Our velocity
measurements complement those obtained by \citet{Kollmeier2009} and
\citet{Sesar2010} who found two groups of stars, one moving towards us
and one away.  We conclude by comparing the observed data to a
theoretical expectation of the stellar wake forming in the halo as the
LMC falls into the Galaxy.

\section{Pisces Plume with {\it Gaia} DR2 and PS1 RR Lyrae}

The analysis presented here is based on the combined sample of RR
Lyrae stars of ab type, with the bulk of the data coming from the {\it
  Gaia} DR2, augmented by the RR Lyrae identified in the Pan-STARRS1
(PS1) survey by \citet{Sesar2017}. More precisely, from the {\it Gaia}
DR2 we select RR Lyrae candidates
\citep[see][]{Clementini2018,Holl2018} with BP/RP excess factor less
than 1.5, located in the areas of the sky with low or moderate dust
reddening i.e. $E(B-V)<0.7$. To clarify, the {\it Gaia} DR2 RR Lyrae
rely on both the results of the general variability analysis
(e.g. \texttt{vari\_classifier\_result} and
\texttt{vari\_time\_series\_statistics} tables) and the SOS (Specific
Object Studies) analysis (\texttt{vari\_rrlyrae} table). In short, our
{\it Gaia} selection is similar to that discussed in
\citet{Iorio2019}. From the PS1 dataset \citet{Sesar2017}, we select
the most likely RR Lyrae candidates by applying a cut on {\it ab
  score} (a statistic akin to the probability of being an RR Lyrae
star) above 0.8. Finally, we exclude objects in the central portion of
the Galactic disk, i.e. those with $|l| < 120^{\circ},
-14^{\circ}<b<12^{\circ}$. The combination of above cuts yields a
sample with a total of $\sim98,000$ RR Lyrae, of which some
$\sim4,000$ are from the PS1.

Figure~\ref{fig:distant} shows the density of GDR2+PS1 RR Lyrae with
Galacto-centric distances between $R=60$ kpc and $R=100$ kpc in
equatorial coordinates. As evidenced in the Figure, the distant halo
of the Milky Way is dominated by three prominent over-densities. The
largest of these, at RA$\sim130^{\circ}$ is associated with the Sgr
stream, more precisely with its trailing tail apo-centre. The red
dashed lines mark the Sgr stream track, as approximated by a great
circle with a pole at (RA, Dec)$=(283.750^{\circ},
-30.483^{\circ})$. The lines correspond to the stream-aligned
latitudes $B_{\odot}=-10^{\circ}$ and $B_{\odot}=10^{\circ}$
\citep[see][for details]{Belokurov2014}. The structure at
RA$=-140^{\circ}$, the so-called Outer Virgo Overdensity (OVO), lies
beyond the apo-centre of the stream's leading tail and may or may not
be connected to the Sgr dwarf disruption
\citep[see][]{Sesar2017,Nina2017}. The lower edge of the third
overdensity, at RA$\sim-10^{\circ}$ appears to overlap with the Sgr
stream track. However, the whole structure extends much further, at
least $\sim20^{\circ}$ out of the Sgr plane. Moreover, there is no
simple way to link this material to the other known portions of the
Sgr stream. Both the observations and the available models agree: the
Sgr stream at this position on the sky is at $D<30$ kpc.

\begin{figure*}
  \centering
  \includegraphics[width=0.97\textwidth]{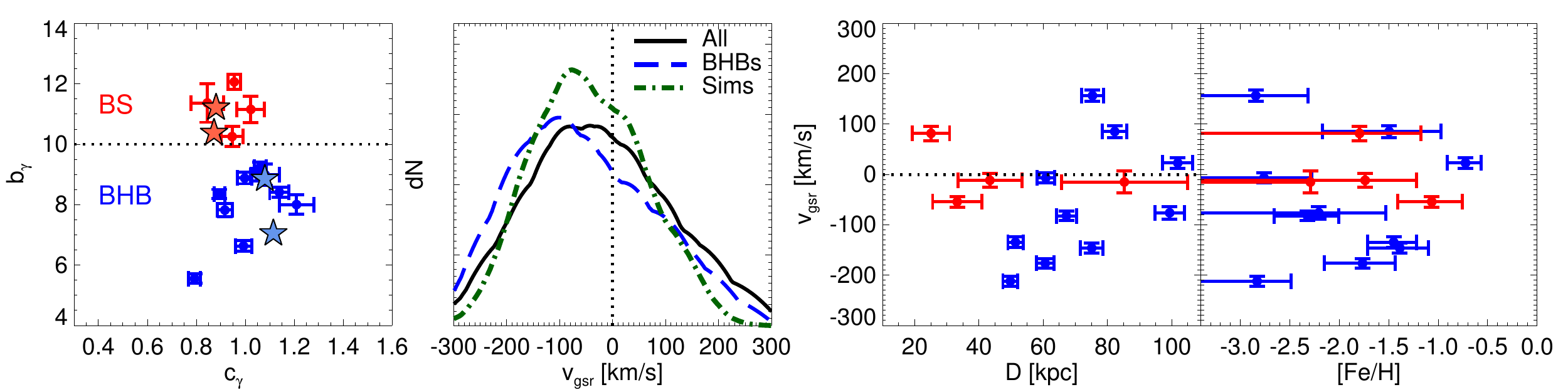}
  \caption[]{{\it Left:} Sersic profile parameters for the H$\gamma$
    Balmer line, $c_\gamma, b_\gamma$. BHB stars are shown in blue and
    BS stars are shown in red. The star symbols indicate the known
    BHB/BS stars from the \cite{xue11} SDSS sample. The BHB and BS
    stars separate at $b_\gamma \sim 10$. {\it Middle-left:} The
    line-of-sight velocity distribution of the A-type stars in
    Galactocentric coordinates. The black line is for all A-type
    stars, the dashed blue line indicates the BHB star distribution,
    and the dot-dashed green line shows the velocity distribution from
    a simulated stellar halo affected by the LMC's in-fall (see
    Section \ref{sec:sims}). The BHB stars have a broad distribution,
    with a bias towards negative radial velocities, in good agreement
    with the simulations. {\it Middle-right:} Line-of-sight velocity
    against distance. The BHB stars in the distance range $40 < D_{\rm
      BHB}/\mathrm{kpc} < 80$ have negative radial velocities. {\it
      Right:} Line-of-sight velocity against metallicity. The stars in
    the VVDS field have similar metallicities to the ``field'' halo
    population.}
   \label{fig:vvds}
\end{figure*}

The structure at RA$\sim-10^{\circ}$ coincides with the previously
reported detections of the so-called Pisces Overdensity
\citep[][]{Sesar2007,Watkins2009}. The distances also broadly agree:
the RR Lyrae displayed here are at D$>60$kpc, while the Pisces
structure is at $\sim80$kpc. Strikingly, as the solid blue lines in
Figure~\ref{fig:distant} indicate, Pisces lies very close to the
equator of the Magellanic Stream (MS). The MS coordinate system is
suggested by \citet{Nidever2008} and is supposed to be aligned with
the Magellanic HI stream. The blue lines correspond to the MS-aligned
latitudes $B_{\rm MS}=\pm15^{\circ}$. Not only is Pisces located at
the point of crossing of the Sgr and Magellanic streams, its shape
appears to be stretched along the MS. The fact that on the sky the
Pisces projected density distribution shows a clear preferred
direction has been noticed before. Namely, \citet{Nie2015} built the
first comprehensive map of Pisces using BHB stars selected from the
deep $u$-band imaging survey SCUSS and found a great circle aligned
with the structure's elongation. The orientation of this great circle
is shown in Figure~\ref{fig:distant} with a black arrow. Clearly, the
Pisces elongation as traced by \citet{Nie2015} deviates only slightly
from the track of the MS. Note that the Pisces morphology shown in
Figure~\ref{fig:distant} may bear some signs of selection biases
inherent to the GDR2 and PS1 datasets. In particular, the GDR2 RR
Lyrae selection at faint magnitudes depends sensitively on the number
of Gaia scans, distributed in a highly non-uniform fashion across the
sky. Reassuringly however, for Dec$>-10^{\circ}$, the shape of the
structure traced with RR Lyrae agrees well with that mapped using the
BHBs, thus at these latitudes the effects of the Gaia scanning law may
be minimal. At lower Dec, we expect a more serious selection bias: the
BHB study of \citet{Nie2015} is limited to Dec$>-10^{\circ}$, while
the PS1 sample does not reach below Dec$=-30^{\circ}$.

The right panels of Figure~\ref{fig:distant} show the galacto-centric
distance distribution of GDR2+PS1 RR Lyrae as a function of the Sgr's
longitude $\tilde{\Lambda}_{\odot}$ for two ranges of Sgr's latitude
$\tilde{B}_{\odot}$. In these panels, the Pisces Over-density appears
as a nearly vertical plume of stars, stretching from $\sim40$ kpc to
beyond 100 kpc at around $\tilde{\Lambda}_{\odot}\sim300^{\circ}$. As
indicated by the red dashed lines, the steep distance gradient
apparent in the RR Lyrae distribution is consistent with that
previously measured using the SCUSS BHBs by \citet{Nie2015}. On
comparing the two right panels of the Figure, it is apparent that the
distribution of the RR Lyrae along the line of sight changes slightly
as a function of $\tilde{B}_{\odot}$.  At high $\tilde{B}_{\odot}$
(second panel) the Pisces Plume is the only significant sub-structure
in the outer MW halo. Within the Sgr orbital plane (third panel), the
dwarf's trailing tail becomes clearly visible, dropping from $r\sim60$
kpc at $\tilde{\Lambda}_{\odot}\sim200^{\circ}$ to $\sim20$ kpc at
$\tilde{\Lambda}_{\odot}\sim300^{\circ}$. Curiously, apart from the
Pisces Plume itself (mostly limited to
$300^{\circ}<\tilde{\Lambda}_{\odot}<320^{\circ}$), there is a narrow
and nearly vertical stream at
$\tilde{\Lambda}_{\odot}\sim330^{\circ}$. Given the position on the
sky and the distance gradient (and as indicated with the small red
filled circles), at least some of this additional signal can be
attributed to the Orphan Stream (OS), which has recently been detected
in this area of the sky \citep[see][]{Koposov2019}. Note that, as
shown by solid black lines in the left panel of
Figure~\ref{fig:distant}, the OS is much narrower than either the Sgr
stream or the Pisces Plume.

\section{Pisces Plume with Subaru HSC and VLT FORS2}

Recently, \citet{Deason2018} used the deep multi-band photometry
acquired as part of the Subaru Hyper Suprime-Cam (HSC) survey to
select candidate BHB and Blue Straggler (BS) stars at large distances
from the Sun. Once completed, the HSC imaging survey will provide one
of the deepest views of a large portion of the Milky Way halo. The
subset of the data available currently is limited to 7 fields with a
combined area of $\sim100$\ deg$^2$. \citet{Deason2018} used a
combination of $griz$ filters to identify BHB and BS stars beyond 50
kpc and as far as 200 kpc. They point out that out of the 7 fields
considered in their analysis, three stand out clearly as they contain
a significantly larger number of distant tracers. Two of these are
projected onto the Sgr stellar stream, the likely source of
contamination. They argue that the third over-dense HSC field, namely
VVDS, lies close to the equator of the MS coordinate system (see
Figure~\ref{fig:distant}) and therefore may be related to the
Magellanic Clouds. As the Figure indicates, the VVDS field also
overlaps with the over-density of PS1+GDR2 RR Lyrae associated with
the Pisces Plume.

\subsection{Spectroscopic Follow-up with VLT/FORS2}

To explore the VVDS overdensity further, we performed spectroscopic
follow-up of BHB candidates in the VVDS field with VLT/FORS2. The
candidates were selected from the HSC data, with $0.0 < griz < 0.07$,
where $griz = i - z - 0.3(g - r) + 0.035$. Our original target list
had $N=19$ candidates, but after inspection of the \textit{Gaia} DR2
astrometry of the relatively bright candidates ($g < 20$), we found
that 2 of these were obvious white dwarfs (with large proper
motions). Our observations were taken over 4 half nights in visitor
mode from 6-10 October 2018 (PI:Deason, ID:0102.B-0029(A)) using the
FORS2 spectrograph. The instrument was used in long slit mode with the
E2V detector, SR collimator, a binning of $2\times2$, and a 1.0 arcsec
slit. We used the 1200B+97 grism, with a dispersion of 0.36 \AA\ per
pixel. The wavelength range provided by this configuration spans
$3600-5110$ \AA\. Our targets spanned a magnitude range of $ 20 < g <
21.5$, and the exposure times ranged from 20 minutes to 2 hours
depending on the apparent magnitude and weather conditions.

Data were taken for 17 targets during our observing run. 15 of these
had good data (S/N $>10$), while 2 faint targets (with $g > 21$) had
insufficient signal-to-noise (S/N $\ll 10$). We also observed 4 bright
BHB/BS stars from the \cite{xue11} catalog. These were used to test
our separation of BHB and BS stars, and evaluate our velocity
precision. The data were reduced using the standard ESO pipelines
using the \textsc{esorex} package. These procedures included
bias-subtraction, flat fielding correction, spectral extraction, sky
correction, and wavelength calibration.

\subsection{Radial velocities and A-type star classification}

\begin{figure*}
  \centering
  \includegraphics[width=0.97\textwidth,trim={0.5cm 0.0cm 1.9cm 0cm},clip]{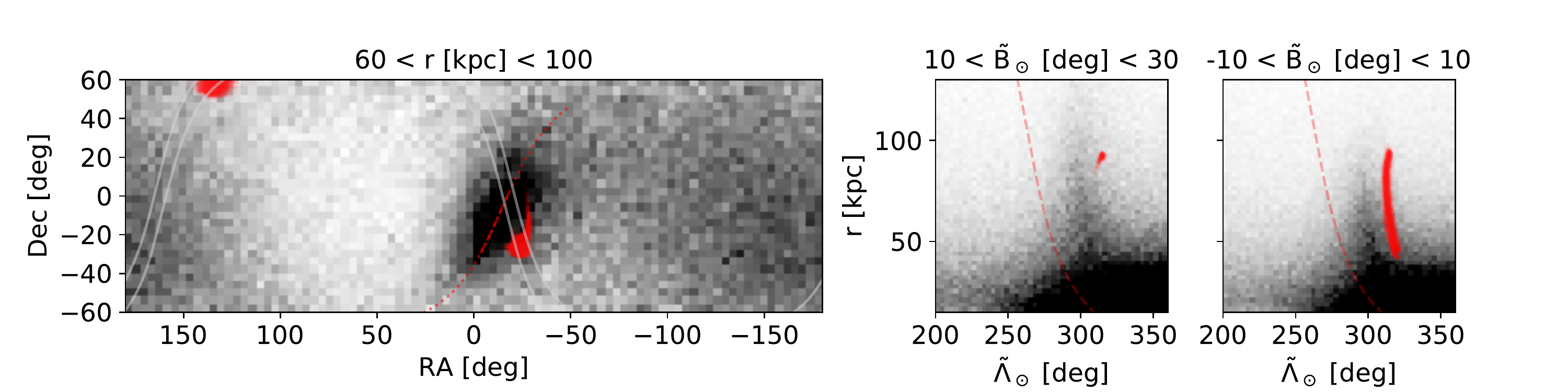}
  \caption[]{Mock observation of simulated stellar halo perturbed by a
    $1.5\times 10^{11} M_\odot$ LMC with same panels as
    Fig. \protect\ref{fig:distant}. \textit{Left:} On-sky density of
    the stellar halo with Galactocentric distances between 60 and 100
    kpc. The dotted red line shows the orbit of the LMC over the past
    5 Gyr while the dashed red line shows the portion of the LMC orbit
    within Galactocentric distances between 60 and 100 kpc. The
    overdensity occurs in precisely the same part of the sky as the
    Pisces overdensity (see Fig. \protect\ref{fig:distant}). The
    overdensity is aligned with the orbit of the LMC in 3D, following
    the overall orbit on the sky as well as matching its extent
    restricted to the same radial range. The grey lines show
    $\pm3^\circ$ from the Orphan stream plane and the red points show
    the best-fit Orphan stream model from
    \protect\cite{orphan_lmc_mass} with the same Galactocentric
    distance cuts.  \textit{Right:} Distribution of mock stellar halo
    stars in heliocentric distance versus Sgr's longitude, $\tilde
    \Lambda_\odot$, restricted to two different ranges of Sgr's
    latitude, $\tilde B_\odot$. The dashed-red line shows the distance
    gradient from \protect\cite{Nie2015} and the red points show the
    best-fit Orphan stream model from
    \protect\cite{orphan_lmc_mass} subject to the same cuts in $\tilde B_\odot$. The overdensity in the wake
    appears for both latitude ranges, extending out to larger
    distances at higher latitudes. For $10^\circ < \tilde{B}_\odot <
    30^\circ$, the wake has a qualitatively similar distance slope and
    location in $\tilde\Lambda_\odot$ to what is observed (see
    Fig. \protect\ref{fig:distant}). At lower latitudes, the wake
    matches the broad component seen in the data but cannot reproduce the
    narrow feature. Instead, this feature can be explained by
    the predicted extension of the Orphan stream. Interestingly, the closest approach between the Orphan stream and the LMC occurs within this extension \protect\citep[see][]{orphan_lmc_mass}, suggesting it will be useful for further constraining the LMC mass.}
   \label{fig:sims}
\end{figure*}

We fit Sersic profiles to the H$\alpha$, H$\gamma$ and H$\beta$ Balmer
lines (see \citealt{navarrete19} eqn. 1), and the radial velocities
are calculated by fitting all 3 lines simultaneously. The models are
convolved with a Gaussian with FWHM=2.88 \AA\ (corresponding to 4
pixels). The errors on the derived velocities include the fitting
errors and the wavelength calibration uncertainty ($\sim 0.15$
pix). An additional source of uncertainty is the offset in the
centering of the star in the slit. Using repeat measurements, and
standard BHB/BS stars, we estimate that this additional error source
can reach up to $\sim 10$ km s$^{-1}$. However, we don't include this
error in our analysis, as this is an upper bound and each star would
have a slightly different offset.

In order to separate BHB and BS stars, we use the Sersic profile of
the H$\gamma$ line, which we found to be the most discerning
(cf. \citealt{clewley02}). In the left hand panel of
Fig. \ref{fig:vvds} we show the $b_\gamma$ and $c_{\gamma}$ Sersic
indices for our A-type stars. The star symbols show the 4 A-type stars
from \cite{xue11}. The BHB and BS stars are clearly separated in this
plane, and the confirmed BHB/BS stars are shown in blue/red. The
velocity distribution of these stars is shown in the middle-left
panel. When converting to Galactocentric velocities, both barycentric
and heliocentric corrections are applied, and we assume the solar
velocity is $(V_x, V_y, V_z )= (11.1, 12.24+235, 7.25)$ km
s$^{-1}$. The velocity distribution of the BHB stars is broad, but
with a bias towards negative velocities: $\left(\sigma(v_{\rm gsr}),
\langle v_{\rm gsr} \rangle\right) = (118^{+35}_{-24},
-57^{+39}_{-39})$ km s$^{-1}$. The middle-right panel shows the
Galactocentric velocity as a function of distance. We assign distances
to the BHB and BS stars using the relations given in
\cite{deason11}. Here, we can again see that the BHB stars in the
radial range 40-80 kpc have a broad velocity distribution, and are
biased towards negative values.
values.

To further investigate these stars in the VVDS field, we estimate
their metallicities ([Fe/H]) using the equivalent width of the
Ca\textsc{ii} K line at $3933$ \AA\ . We adopt the relations derived
in \cite{navarrete19}. For consistency, we check that our estimated
metallicities agree with the SDSS SEGUE metallicities for the known
BHB/BS stars: we find that they all agree within $1\sigma$. The
resulting [Fe/H] values are shown in the right hand panel of
Fig. \ref{fig:vvds} against velocity. Here, we can see no obvious
chemical signature and the BHB stars appear to have metallicities
typical of ``field'' halo stars, with [Fe/H] $\sim -2.0$. In summary,
the excess of BHB stars in the VVDS field have a broad velocity
distribution, with a ($\sim$$1.5\sigma$ significance) bias towards
negative radial velocities. Moreover, these stars have metallicities
consistent with the field halo population.

\section{Discussion and Conclusions}

\subsection{Comparison to simulations}
\label{sec:sims}
Although the plume in Figure \ref{fig:distant} appears somewhat
stream-like, the radial velocities in Figure \ref{fig:vvds} do not
reveal any obvious cold components. Instead, they show a dispersion
consistent with that of the MW's stellar halo
\citep[e.g.][]{bird_anisotropy, Lancaster2019} with a slight bias
towards negative line-of-sight velocities. Interestingly, such a shift
in the stellar halo's velocity was predicted in \cite{orphan_lmc_mass}
who argued that the LMC's infall should induce a substantial reflex
motion in the inner regions of the Milky Way relative to its
outskirts. This effect would push the mean radial velocity of the
stellar halo in the Pisces region to be negative. This effect was also
seen in \cite{lmc_wake} who simulated the LMC's in-fall and found that
it should also induce a substantial wake, i.e. overdensity, in the
Milky Way's stellar halo.

In order to test this prediction, we evolve the Milky Way in the
presence of various mass LMCs. This is done using a similar setup as
\cite{orphan_lmc_mass}. Crucially, the Milky Way is treated as a
particle sourcing the \texttt{MWPotential2014} from \cite{bovy} where
we have replaced the bulge with a Hernquist bulge with a mass of
$5\times10^9 M_\odot$ and a scale radius of $500$ pc for computational
efficiency. This allows the Milky Way to respond to the LMC's infall
which is essential for correctly capturing the LMC's full effect. The LMC
is modelled with a Hernquist profile with a mass of
$[2,5,10,15,20,25]\times10^{10} M_\odot$. For each LMC mass, we fix
the scale radius by requiring the circular velocity at 8.7 kpc matches
the observed value of 91.7 km/s from \cite{lmc_circular_vel}. The
stellar halo is set up in equilibrium using \textsc{agama}
\citep{agama}. Namely, we use the \texttt{DoublePowerLaw} distribution
function which is similar to the two-power models described in Sec.
3 of \cite{posti_etal_2015}. We use \texttt{norm=1.5e10},
\texttt{j0=500}, \texttt{slopeIn=0}, \texttt{slopeOut=3.5},
\texttt{coefJrOut=0.75}, \texttt{coefJzOut=1.125},
\texttt{jcutoff=1e5}, \texttt{cutoffStrength=2}. This produces a
nearly spherical stellar halo with an outer power-law slope of -3.2 beyond
$\sim20$ kpc, and an almost constant anisotropy of 0.475 between
10-500 kpc. This broadly resembles the observed stellar halo's
properties, albeit with a slightly shallower density fall off
\citep[e.g.][]{deason11} and a constant instead of falling anisotropy
\citep[e.g.][]{bird_anisotropy}. We sample this stellar halo with
$10^7$ particles.

In each simulation, the Milky Way is initialized at the origin and the
LMC is initialized with its observed radial velocity
\citep{van_der_marel+2002}, proper motion \citep{Kallivayalil13}, and
distance \citep{lmc_dist}. They are rewound for 5 Gyr in the presence
of each other, including dynamical friction from
\cite{Jethwa2016}, using a kick-drift-kick integrator. At that point,
the stellar halo is injected around the Milky Way as tracer particles
and the system is evolved to the present. Mock observations are
made on the final snapshot from the location of the Sun, 8.122
kpc from the Milky Way center \citep{gravity_R0}. This stellar halo
has all of the same qualitative features as those reported in
\cite{lmc_wake}. This process is repeated for all 6 LMC masses. In
order to test the stability of our initial conditions, and to compare
with the LMC's effect, we also consider a fiducial case where the LMC
is not included. In this case, the stellar halo's properties are
stable and do not evolve significantly over 5 Gyr.

Figure \ref{fig:sims} shows mock observations of the simulated stellar
halo with the effect of a $1.5\times 10^{11} M_\odot$ LMC \citep[which
  roughly matches the LMC mass measured in][]{orphan_lmc_mass} using
the same cuts as in Figure~\ref{fig:distant}.  Interestingly, there is
a sharp plume-like feature around RA $\sim-20^\circ$ ($\tilde
\Lambda_\odot \sim 300^\circ$) which matches the observed Pisces Plume
shown in Figure~\ref{fig:distant}. As expected, this overdensity is
aligned with the past orbit of the LMC. The dot-dashed green line in
the middle-left panel of Fig. \ref{fig:vvds} shows the distribution of
radial velocities in the VVDS footprint, which we approximate with the
rectangle, $331^\circ < {\rm RA} < 341^\circ$, $-0.5^\circ < {\rm dec}
< 2^\circ$, and distances, $40 \, {\rm kpc} < D < 110 \, {\rm
  kpc}$. This agrees well with the observed velocity distribution. As
in the observations, there is a broad distribution with a shift
towards negative velocities. This shift increases with increasing LMC
mass. For example, in order to match the observed shift of $\sim -57$
km/s, an LMC mass of $\sim 2\times 10^{11} M_\odot$ is required. Figure \ref{fig:sims} also shows the best-fit Orphan stream from \cite{orphan_lmc_mass} in (red points) which matches the thin, almost vertical, stream visible in the right panel of Figure~\ref{fig:distant}. 

\subsection{Conclusions}

We have used a sample of the RR Lyrae detected in the Gaia DR2 and PS1
data to trace a long and a nearly radial stellar stream, a portion of
which had been identified earlier as the Pisces Overdensity. Given the
stream's stretched appearance in 3D, we have dubbed it the Pisces
Plume. The Plume lies at the cross-section of the equators of the Sgr
and Magellanic Stream coordinate systems. If the Plume is of the Sgr
origin then it must be connected to a previously unseen wrap of the
Sgr stream, significantly out of the current Sgr plane. However, the
alignment of the elongated shape of the Pisces Plume on the sky with
the MS equator suggests a possible Magellanic origin.

We have collected VLT FORS2 follow-up spectroscopy of 17 candidate BHB
stars detected on the edge of Pisces by \citet{Deason2018} using deep
Subaru imaging. Interestingly, radial velocity measurements of
candidate Pisces members show a broad dispersion, consistent with the
stellar halo, with a shift towards negative values. Guided by previous
predictions of the LMC's effect on the Milky Way's stellar halo,
\citep[e.g.][]{orphan_lmc_mass,lmc_wake}, we simulated the LMC's
infall and found that it reproduces well the Pisces Plume's 3D
morphology and its radial velocity distribution.  This overdensity is
the first direct evidence of dynamical friction in action: as the LMC
passes through the Milky Way, its gravitational effect induces a wake
of stars and dark matter behind it. Since the wake is aligned with the
LMC's past orbit, the Pisces Plume offers a way to measure the
dynamical friction exerted on the LMC by the Milky Way. This will shed
light on the nature of dark matter since dynamical friction depends on
its microscopic properties \citep[e.g.][]{hui+2017}. Finally, this
interpretation of the Pisces Plume also suggests that the outskirts of
the Milky Way are out of equilibrium due to the effect of the LMC on
our Galaxy. Indeed, the results of \cite{lmc_wake} suggest that we
will detect significant velocity offsets over the entire sky, ushering
in a new era of disequilibrium modelling.

\section*{Acknowledgments}

The research leading to these results has received funding from the
European Research Council under the European Union's Seventh Framework
Programme (FP/2007-2013) / ERC Grant Agreement n. 308024. V.B.,
D.E. and S.K. acknowledge financial support from the ERC. A.D. is
supported by a Royal Society University Research Fellowship. A.D. also
acknowledges support from the STFC grant ST/P000451/1. SK is partially
supported by NSF grant AST-1813881. JAC-B acknowledges financial
support to CAS-CONICYT 17003. MCS acknowledges financial support from
the National Key Basic Research and Development Program of China
(No. 2018YFA0404501) and NSFC grant 11673083. PJ acknowledges funding
from the European Research Council (ERC) under the European Union’s
Horizon 2020 research and innovation programme under grant agreement
No 724857 (Consolidator Grant ArcheoDyn) This research made use of
data from the European Space Agency mission Gaia
(http://www.cosmos.esa.int/gaia), processed by the Gaia Data
Processing and Analysis Consortium (DPAC,
http://www.cosmos.esa.int/web/gaia/dpac/consortium). Funding for the
DPAC has been provided by national institutions, in particular the
institutions participating in the Gaia Multilateral Agreement. This
paper made used of the Whole Sky Database (wsdb) created by Sergey
Koposov and maintained at the Institute of Astronomy, Cambridge with
financial support from the Science \& Technology Facilities Council
(STFC) and the European Research Council (ERC).

\bibliography{references}

\label{lastpage}

\end{document}